**Tunable Anomalous Hall Effect in a Kagomé Ferromagnetic Weyl Semimetal**

*Samuel E. Pate, Bin Wang, Yang Zhang, Bing Shen, Enke Liu, Ivar Martin, J. Samuel Jiang, Xiuquan Zhou, Duck Young Chung, Mercouri G. Kanatzidis, Ulrich Welp, Wai-Kwong Kwok, and Zhi-Li Xiao*[*]

Samuel E. Pate, Ivar Martin, J. Samuel Jiang, Xiuquan Zhou, Duck Young Chung, Mercouri G. Kanatzidis, Ulrich Welp, Wai-Kwong Kwok, and Zhi-Li Xiao
Materials Science Division, Argonne National Laboratory; Argonne, 60439, United States
E-mail: xiao@anl.gov

Samuel E. Pate, Zhi-Li Xiao
Department of Physics, Northern Illinois University; DeKalb, 60115, United States

Bin Wang, Bing Shen
School of Physics, Sun Yat-sen University; Guangzhou, 510275, China

Yang Zhang
Department of Physics, University of Tennessee; Knoxville, 37996, United States
Min H. Kao Department of Electrical Engineering and Computer Science, University of Tennessee, Knoxville, 37996, United States

Enke Liu
Institute of Physics, Chinese Academy of Sciences; Beijing, 100190, China

Mercouri G. Kanatzidis
Department of Chemistry, Northwestern University; Evanston, 60208, United States

Emerging from the intricate interplay of topology and magnetism, the giant anomalous Hall effect (AHE) is the most known topological property of the recently discovered kagomé ferromagnetic Weyl semimetal $Co_3Sn_2S_2$ with the magnetic Co atoms arranged on a kagomé lattice. Here we report that the AHE in $Co_3Sn_2S_2$ can be fine-tuned by an applied magnetic field orientated within ~2° of the kagomé plane, while beyond this regime, it stays unchanged. Particularly, it can vanish in magnetic fields parallel to the kagomé plane and even decrease in magnetic fields collinear with the spin direction. This tunable AHE can be attributed to local spin switching enabled by the geometrical frustration of the magnetic kagomé lattice, revealing that spins in a kagomé ferromagnet change their switching behavior as the magnetic field approaches the kagomé plane. Our results also suggest a versatile way to tune the properties of a kagomé magnet.

## 1. Introduction

Weyl semimetals are emergent materials hosting relativistic Weyl fermions when inversion symmetry or time-reversal symmetry is broken.[1-3] While the inversion symmetry-broken Weyl semimetals were discovered more than a decade ago and have been extensively studied,[1] the time-reversal symmetry-broken ones were experimentally confirmed only recently, starting with the ferromagnetic $Co_3Sn_2S_2$, as revealed by the giant anomalous Hall effects (AHEs)[4,5] and characteristic surface Fermi-arcs and linear bulk band dispersions near the Weyl points.[3,6] The interplay of topology and magnetism leads to abundant fascinating phenomena such as anomalous Nernst effect,[7] negative magnetoresistance due to a chiral anomaly,[4] unusually large magneto-optic effects,[8] and the largest Hall angle reported.[9]

The crystal structure of $Co_3Sn_2S_2$ comprises of $Co_3Sn$ layers in the *ab* plane separated by $SnS_2$ blocks (Figure S1A),[10] with the magnetic Co atoms arranged on a kagomé lattice (**Figure 1A**, Figure S1B),[11,12] a hallmark of other newly discovered topological kagomé magnets and superconductors with exotic properties.[13] Its ferromagnetic (FM) state shows strong anisotropy, with the magnetization easy-axis along the *c*-axis (Figure 1B, Figure S1C&D).[14] The combination of the kagomé lattice structure with long-range out-of-plane FM order makes $Co_3Sn_2S_2$ an excellent candidate for observing exotic topological quantum states such as the quantum anomalous Hall state in the two-dimensional limit.[4,15] As revealed in both muon spin-rotation (μSR)[16] and neutron scattering experiments,[17] the kagomé lattice could enable the occurrence of frustrated in-plane anti-ferromagnetic (AFM) structures (Figure 1C and Figure S1E&F) competing with the out-of-plane FM order at elevated temperatures. While the existence of the in-plane AFM phases in $Co_3Sn_2S_2$ has been widely debated,[2,11,12,18-20] it seems to be the most plausible origin for the observed exchange-bias effects[21,22] that occur typically only in an FM/AFM bilayer structure.

Here we report unusual spin switching behavior enabled by the kagomé lattice in this FM Weyl semimetal. We find that the AHE arising from the FM order in $Co_3Sn_2S_2$ can be fine-tuned by applying a magnetic field orientated within ~2° of the kagomé plane while beyond this regime the behavior follows that of a conventional ferromagnet. Remarkably, the AHE can suddenly vanish as the in-plane magnetic field reaches a critical value. The amplitude of the AHE can also abruptly decrease at magnetic fields having components even in the same direction as the spins, where no conventional spin flip is expected. The observed AHE behavior can be potentially attributed to local spin switching induced by local lattice instability enabled by the geometrical frustration of the kagomé lattice,[18] which converts an ordered FM structure into a mixture of FM domains and in-plane AFM orders. Our results reveal a remarkable interaction of the external magnetic field and the magnetic kagomé lattice, resulting in distinctive switching behavior of spins in a kagomé magnet. They also shed light on the existence of the highly debated AFM phase in $Co_3Sn_2S_2$,[2,11,12,16-20] which can account for the observed novel phenomena such as bow-tie-like hysteresis loops[21] and exchange bias effects.[21,22]

## 2. Results and Discussion

We conducted both transport and magnetization measurements on single crystals grown by flux methods.[23] More experimental details including the crystal growth and measurement systems can be found in the supplement. Figure S3 and its caption describe the procedures for determining the *ab* plane for precise applied magnetic field orientation. Figure S4 shows a typical zero-field cooling resistance versus temperature curves, revealing a Curie temperature of $T_c \approx 173$ K, consistent with those in the literature.[4,5]

The FM order in $Co_3Sn_2S_2$ is believed to be important in the breaking of time reversal symmetry, which induces the Weyl states.[4,16,17] Current AHE experiments are typically conducted with magnetic fields along the *c*-axis. Figure 1D and Figure S5A present the Hall resistance versus magnetic field ($R_{xy} \sim H$) loops of Samples R1 and R2 at $T = 3$ K for $H \parallel c$, respectively. They show the typical spin-flip governed behavior of a ferromagnet. That is, all spins flip when the magnetic field in the opposite direction of the spins reaches the coercive value, resulting in an instantaneous sign change in the $R_{xy}$ while retaining its amplitude.[24,25] On the other hand, canting the magnetization away from the *c*-axis is predicted to change the locations of the Weyl nodes, affecting the topological properties such as AHEs and Nernst effects.[26] It is also expected to make the FM order unstable.[27] Thus, it is interesting to investigate possible effects of field orientation on topological properties. In fact, recent experiments reveal that magnetic phases in $Co_3Sn_2S_2$ can depend on the field orientation.[28]

To further explore the canting effects in the FM state or more generally the response of the FM structure to a tilted external magnetic field, we first create it by applying a field of $H = -9$ T along the *c*-axis to align all the spins in the same direction. After decreasing the field to zero at which the AHE is observed, we apply the magnetic field at various orientations while recording $R_{xy}$ (Figure S6). As demonstrated in Figure 1E and Figure S5B for magnetic fields orientated not too close to the kagomé plane, the FM structure in $Co_3Sn_2S_2$ behaves the same way as that of a conventional ferromagnet, where flips of all the spins occur and the $R_{xy}$ changes sign at the coercive field.[24,25] However, an unexpected change occurs in $R_{xy}$ when the magnetic field is in the kagomé plane, as presented in Figure 1F and Figure S5C, which show a sudden change of $R_{xy}$ to a value near 0 as the magnetic field increases to a switching field $H_s$. These results are repeatable (Figure S7A) in consideration of experimental accuracies such as rotator resolution and reproducibility as well as the nature of the mechanism inducing the sudden change in $R_{xy}$, as discussed below. The new state is also stable with field excursion (Figure S7B&C).

Since the in-plane saturation field is as high as $H_{ab}^S = 23$ T,[14] the sudden occurrence of $R_{xy}^S \approx 0$ at a much lower field $H_s$ is not due to the collective flop of the spins from the $c$-axis to the $ab$ plane. While the annihilation of Weyl nodes has only been observed across the Curie temperature,[29] it could lead to $R_{xy}^S \approx 0$ at $H_s$ if associated with a canted magnetic field.

To clarify these two scenarios, we conducted magnetization measurements on the FM structure driven by an in-plane field (see Figure S8 for procedures). The insets of Figures 2A and 2B show that the in-plane magnetization of the $c$-axis FM structure of sample M1 increases almost linearly with the in-plane fields up to $H = 7$ T. We expected to see the in-plane magnetization to vanish if all the spins in the FM structure are perfectly aligned along the $c$-axis or be the $c$-axis saturation value if all spins are parallel to the in-plane field. The smooth increase of the in-plane magnetization as shown in the insets of Figures 2A and 2B rules out the possibility of a sudden flop of all the spins from the $c$-axis to the $ab$ plane. On the other hand, it could be induced by spin canting away from the $c$-axis, which becomes more pronounced at higher in-plane fields. The main panel of **Figure 2A** shows the magnetizations measured along the $c$-axis after the FM structure was driven by in-plane fields up to $H = 3$ T. It indicates that the FM structure is still intact. In contrast, results in the main panel of Figure 2B reveal that in-plane fields up to $H = 7$ T can eradicate the FM structure, as shown by the rapid decrease in the $c$-axis magnetization at zero field, followed by the saturation behavior with increasing field. More measurements following the same procedures with in-plane field sweeping up to intermediate values, as shown in the main panels of Figures S9A and S9B, enable to determine the lowest in-plane field of 3.37 T for the uniform FM structure to be destroyed. The resemblance of the sudden changes in Figure 2C to those in Figure 1F and Figure S5C suggests that the observed vanishing of $R_{xy}$ at $H_s$ is caused by the destruction of uniform FM order, highlighting again the crucial role of FM order on the occurrence of AHE.

To gain further insights on the field-induced suppression of the FM order, we measured $R_{xy} \sim H$ curves at more angles and in both positive and negative fields to the same FM structure, with emphasis on magnetic fields orientated near the kagomé plane. The results are presented in **Figure 3**. Consistent with those in Figure 1E and Figure S5B, $R_{xy}$ shows a sudden sign change due to conventional spin-flips for both directions of the magnetic field as long as they are not aligned too close to the kagomé plane. As presented in **Figure 4A**, the switching fields can be described with the Kondorsky relation $H_s(\theta) = H_s(0°)/|\cos\theta|$,[30,31] though deviations can be seen as the field approaches the kagomé plane. Figure 4B shows that the corresponding $R_{xy}^S$ stays nearly the same. These results can be accounted for with spin-flips dominated by the field-induced nucleation and motion of domain walls that separate regions of opposite uniform magnetizations.[32]

However, as the field is tilted very close to the kagomé plane, we saw unusual changes in the $R_{xy}$. The data in Figure 3 for $\theta = 90.1°$ in positive fields and $89.9°$ in negative fields confirm the $R_{xy}^S \approx 0$ state seen in Figure 1F and Figure S5C. Besides, we found that $R_{xy}^S$ can have other values, as exhibited by the $R_{xy}^S \sim \theta$ curves in Figures 4B and 4D. That is, the AHE in $Co_3Sn_2S_2$ can be fine-tuned with a magnetic field near the kagomé plane. Figure 4 further shows that the FM structures prepared using $H = \pm 9$ T along both directions of the $c$-axis behave similarly.

The continuous change of $R_{xy}^S$ with angle underlines the mechanism of field-induced spin instability in destroying the FM order. In fact, abrupt changes at magnetic fields having components even in the same direction of the spins (e.g., at $\theta = 90.3°$ for positive fields and 89.5° for negative fields in Figure 3) are difficult to reconcile with conventional spin flips. Field-induced spin-instability is also the only reasonable origin for the similar changes to an FM structure induced by magnetic fields in opposite directions. As discussed above, spin canting could induce spin instability.[27] However, the canting angle $\theta_{cs} = \text{asin}(H_s/H_{ab}^S) < 8°$ at $H_s$ is much smaller than the predicted critical angle of 26°.[27] Such an instability was expected to induce complete destruction of the FM structure, resulting in $R_{xy}^S \approx 0$. This is inconsistent with the observed tunability of $R_{xy}^S$, which indicate possible partial destruction of the FM structure, i.e., uncompensated FM domains can exist in the states of $R_{xy}^S \neq 0$.

Very recently, neutron scattering experiments[18] revealed that the geometrical frustration in the Co kagomé lattice can cause a local lattice instability, with a symmetry change from hexagonal $R\bar{3}m$ to distorted monoclinic $Cm$. Density functional theory (DFT) calculations show that this lattice instability causes the local ferromagnetic moments to rotate ~20°. More importantly, DFT data further show that in the monoclinic lattice the energies of the in-plane magnetic states are only slightly higher than those of the out-of-plane ones. That is, a local lattice instability can lead to the occurrence of in-plane magnetic orders, possibly explaining the coexistence of FM and AFM phases[16] observed in μSR measurements or the glass phases[21] proposed to understand the bow-tie-like hysteresis loops occurring in the magnetic field aligned along the $c$-axis as the temperature is swept up to the Curie temperature[21] or at low temperatures when the magnetic field is rotated from the $c$-axis toward the $ab$ plane.[28]

While the local lattice instability revealed in neutron scattering[18] is temperature driven, we hypothesize that it can also be induced by a magnetic field, enabling an instability at lower temperatures. The accompanying local ferromagnetic instability could result in the observed unusual $R_{xy}$ behavior. The application of a magnetic field of $H = \pm 9$ T along the $c$-axis stabilizes a perfect FM structure. The interaction of the Co magnetic moment with a magnetic field in or near the plane distorts Co kagomé lattice thus could cause a local lattice instability at $H_s$. The associated local ferromagnetic instability then transforms the ordered FM structure into a mixture of FM domains with local lattices of $R\bar{3}m$ and $Cm$ symmetries as well as in-plane magnetic orders, probably AFM structures with a local lattice of $Cm$ symmetry. The competition of the FM domains and AFM domains of various volumes and directions determine the value of the $R_{xy}^S$, leading to its tunability. In fact, the field dependence of $R_{xy}$ in Figure 3 with fields $H > H_s$ on both sides of the plane, i.e., at $\theta \neq 90°$, can only be accounted for with such an instability while excluding other mechanisms such as spin-canting induced instability which leads to the complete destruction of the FM order, i.e., $R_{xy} = 0$.[27] Figure 4 also shows that the $H_s \sim \theta$ curves for the same magnetic torque directions reflect each other while exhibiting significant deviations for opposite magnetic torque directions, as expected from a field-induced local lattice instability, since the exchange interactions from the neighbors of a spin in a kagome lattice are not antisymmetric with respect to the magnetic torques induced by opposite magnetic fields (except when the torques are perpendicular to the reflection axes, which is rarely achievable in experiments). On the other hand, the $R_{xy}^S \sim \theta$ curves for the FM structures of the same magnetization direction in Figures 4B and 4D nearly overlap each other with a rotation of 180°, demonstrating again that AHEs are governed by the FM orders. The repeatable formation of a perfect FM order by applying a field of $H = \pm9$ T along the $c$-axis and the field-driven recovery of the FM order after the instability exhibited in Figure 3 at $H > H_s$ and $\theta \neq 90°$ imply that an external magnetic field can also clean up the distorted local lattice of $Cm$ symmetry, restoring $R\bar{3}m$ symmetry of the lattice. In fact, the regime ($< 2°$) near the kagomé plane where $R_{xy}$ can be tuned seems to be limited by the saturation field $H_c^S$ in the $c$-axis direction. That is, the field-induced local FM instability may still occur outside this angle regime but the accompanying mixed phases will be cured since the $c$-axis component ($H_s sin2°$) of the field is larger than $H_c^s$ ($\approx 0.1$ T).[14] While verifications using techniques such as neutron scattering[18] are needed, the creation and removal of distorted local lattices by applying a magnetic field suggested by our results constitute a versatile way to tune the properties of a kagomé magnet.

While data presented in Figures 1-4 were all taken at $T = 3$ K, we did carry out experiments at high temperatures as long as an FM structure at zero field can be prepared. Figure S10 shows results obtained at $T = 80$ K, unveiling the same unusual $R_{xy}$ behaviors as those at $T = 3$K (Figure 3). We also present in Figure S11 the temperature dependence of the switching field $H_s$ for $H \parallel ab$ and $H \parallel c$ as well as their ratios. Clearly, $H_s$ decreases with increasing temperature for both $H \parallel ab$ and $H \parallel c$ while we could obtain $H_s^{ab}$ (= 0.46T) in the $ab$ plane up to $T = 150$ K at which $H_s^c$ (= 4 mT) for $H \parallel c$ is close to zero. Notably, the ratio $H_s^{ab}/H_s^c$ changing from 24 at $T = 130$ K to 115 at $T = 150$ K increases much faster with temperature at $T > 130$ K, where FM/AFM coexistence[16] and bow-tie-like hysteresis loops at $H \parallel c$ [21] were reported and accounted for with local FM instability.[18] In fact, the fast increase of the ratio can also be explained with local FM instability, because the $H_s^c$ at $H \parallel c$ decreases faster with increasing temperature due to temperature-enhanced local FM instability while the $H_s^{ab}$ for $H \parallel ab$ dominated by the field-induced local FM instability is less sensitive to the temperature.

In this high temperature regime we also observed bow-tie-like hysteresis loops at $H \parallel c$, as shown in Figure S11, indicating that they arise from the mixed phases produced by the temperature-driven local FM instability. While the local FM instability for $H \parallel ab$ is driven by the applied magnetic field instead of temperature, it should result in a similar mixture of various magnetic phases. Thus, we expect to see bow-tie-like hysteresis loops in the FM structure after the occurrence of the field-driven local FM instability. As presented in **Figure 5** for $T = 3$ K and Figure S13 for $T = 100$ K, the $R_{xy}$ hysteresis loops taken at $H \parallel c$ after the FM structure had been driven by an in-plane field $H_{ab} > H_s$ indeed show bow-tie-like features (see Figure S12 for measurement procedures). These results reveal that bow-tie-like features of the $R_{xy}$ hysteresis loop are hallmarks of a local FM instability driven either by temperature or an applied magnetic field. They provide strong evidence for the existence of the hypothesized field-induced local FM instability and also support the understanding[18] on the origin of the FM/AFM coexistence[16] and the bow-tie-like hysteresis loops at $H \parallel c$ [21] at elevated temperatures.

## 3. Conclusion

We conducted angle dependent Hall resistance measurements on kagomé ferromagnetic Weyl semimetal $Co_3Sn_2S_2$, which reveal that the spins change their switching behavior as the external magnetic field approaches the kagomé plane. Away from the kagomé plane the spins driven by a magnetic field with a component in the opposite direction flip globally, resulting in a sudden sign change in the anomalous Hall effect (AHE) while retaining its amplitude. When the applied magnetic field is oriented within ~2° of the kagomé plane local spin switching occurs, yielding fine-tune of the AHE. Particularly, it can vanish in magnetic fields parallel to the kagomé plane and even decrease in magnetic fields collinear with the spin direction. We attribute the observed tunable AHE to the occurrence of local ferromagnetic instability caused by distorted local lattices induced by the applied magnetic field due to its interaction with the frustrated magnetic kagomé lattice.

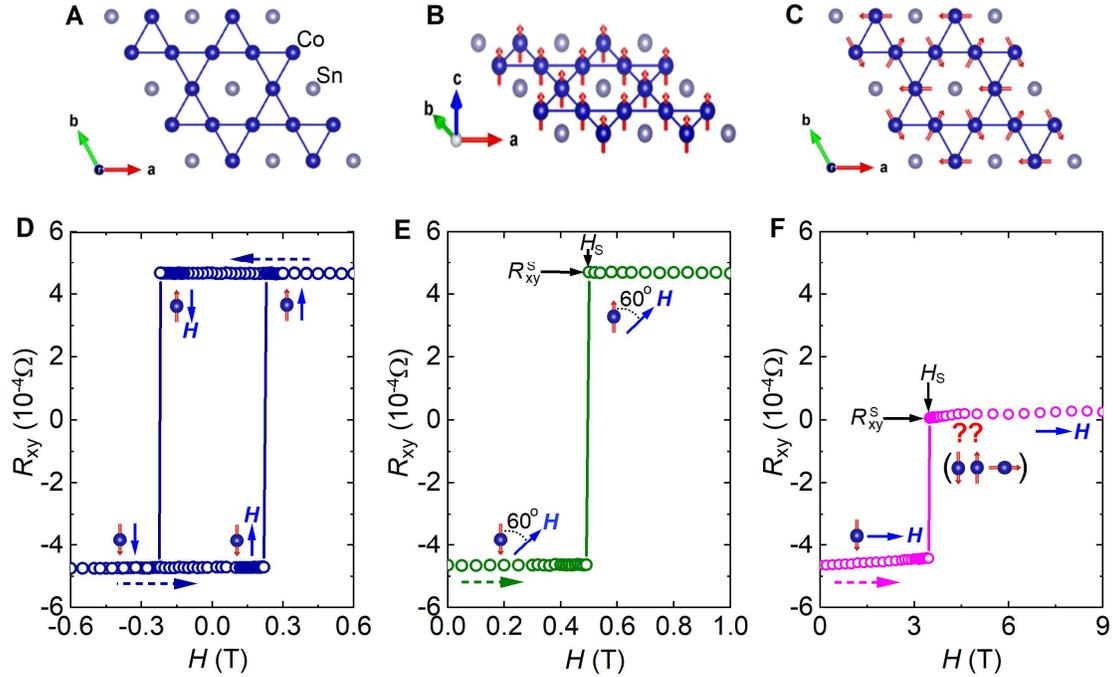

**Figure 1. Hall responses of a Co₃Sn₂S₂ crystal.** (**A**) Structure of the Co₃Sn layer containing the Co kagomé lattice. (**B**) Schematic showing one of the FM states with spins along the positive direction of the *c*-axis. (**C**) Schematic showing one of the AFM states with spins lying in the plane. (**D**) Hysteresis loop of the Hall resistance versus the magnetic field at $H \parallel c$ ($\theta = 0°$). (**E**) and (**F**) Hall responses of an FM structure to magnetic fields orientated at $\theta = 60°$ and $90°$ (in-plane), respectively. The FM structure was prepared by sweeping the field up to $H = -9$ T along the *c*-axis and sweeping back to zero. $\theta$ is the angle between the positive direction of the *c*-axis and the field vector pointing in the positive direction of the magnetic field. $H_s$ and $R_{xy}^S$ in (E) and (F) are the field and Hall resistance right after the spin switching. More details are presented in Figures S3 and S6 and their captions. Data were taken from Sample R1 at $T = 3$ K. The magnetic field rotates in the plane perpendicular to the current direction.

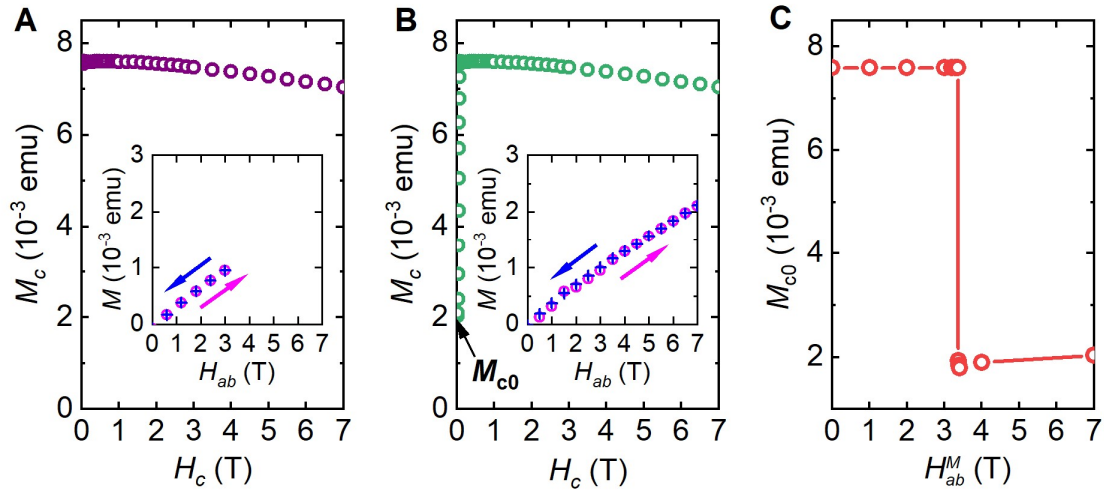

**Figure 2. Magnetization of the FM structure being and after driven by an in-plane field.** (**A**) and (**B**) $M_c$ versus $H_c$ curves of the FM structure after driven by an in-plane field $H_{ab}$ up to $H_{ab}^M = 3$ T and 7 T, respectively, where $M_c$ is the magnetization measured with the magnetic field $H_c$ aligned along the $c$-axis. Their insets represent the in-plane magnetization $M_{ab}$ for the FM order driven by the in-plane field $H_{ab}$. The FM structure was prepared by sweeping the field up to $H = 7$ T along the $c$-axis and sweeping back to zero. Detailed procedures are presented in Figure S8 and its caption. (**C**) Relationship between the zero-field $c$-axis magnetization $M_{c0}$ and the $H_{ab}^M$. Definition of $M_{c0}$ is given in (B). Data were taken from Sample M1 at $T = 3$ K.

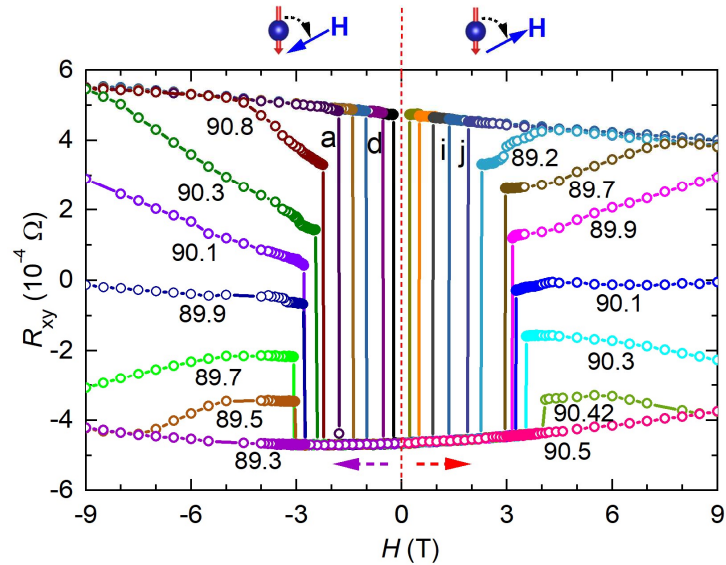

**Figure 3. Hall responses to both positive and negative fields at various angles.** The FM structure with spins pointing to the negative direction of the *c*-axis was prepared by sweeping the field up to *H* = -9 T along the *c*-axis and sweeping back to zero. The red and purple dashed arrows indicate the sweeping directions of the magnetic field for measuring $R_{xy}$. Numbers represent the angles at which the $R_{xy} \sim H$ curves were taken. The angles for other ten curves (from curve a on the left to curve j on the right) are 92°, 93.8°, 98°, 110°, 180°, 0°, 60°, 75°, 82° and 87.8°. The schematics on top of the main panel show the directions of the spins in the FM structure and the magnetic field as well as the definition of the angle. Data were taken from Sample R1 at *T* = 3 K. The magnetic field rotates in the plane perpendicular to the current direction.

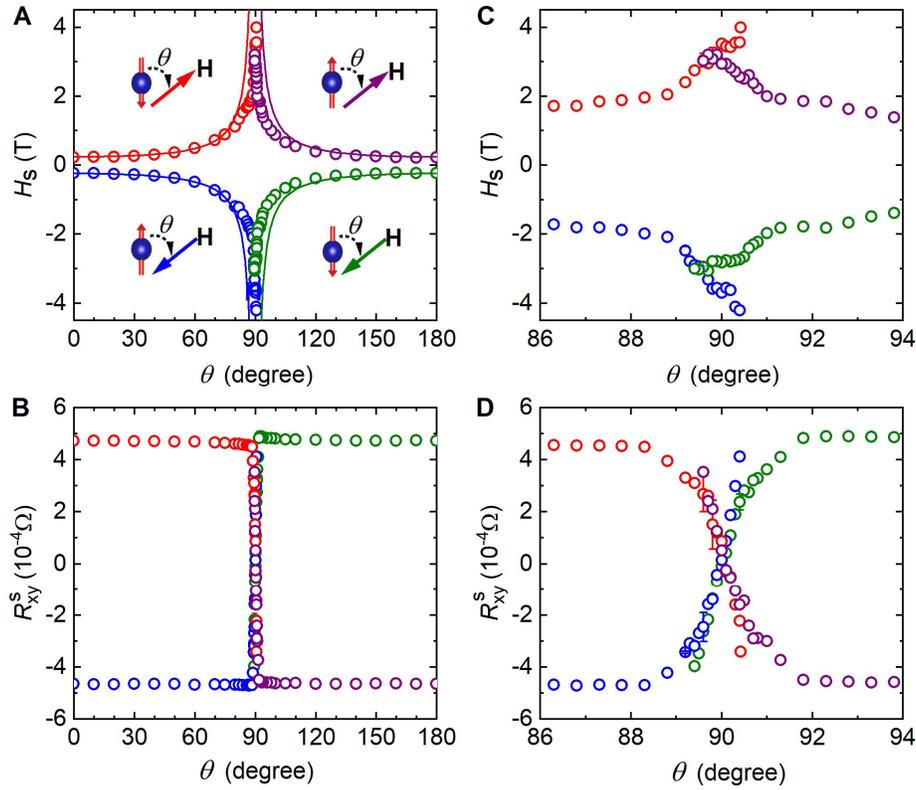

**Figure 4. Angle dependence of the field $H_s$ and Hall resistance $R_{xy}^S$ right after the spin switching.** (**A**) Angle dependence of $H_s$. (**B**) Angle dependence of $R_{xy}^S$. (**C**) and (**D**) Expanded views of (A) and (B) around $\theta = 90°$, respectively. The schematics in (A) show the directions of the spins in the FM structures and the applied magnetic field as well as the definition of the angle $\theta$. Solid lines in (A) represent the Kondorsky relation $H_s(\theta) = H_s(0°)/|\cos\theta|$. Data were obtained from measurements similar to those presented in Figure 3.

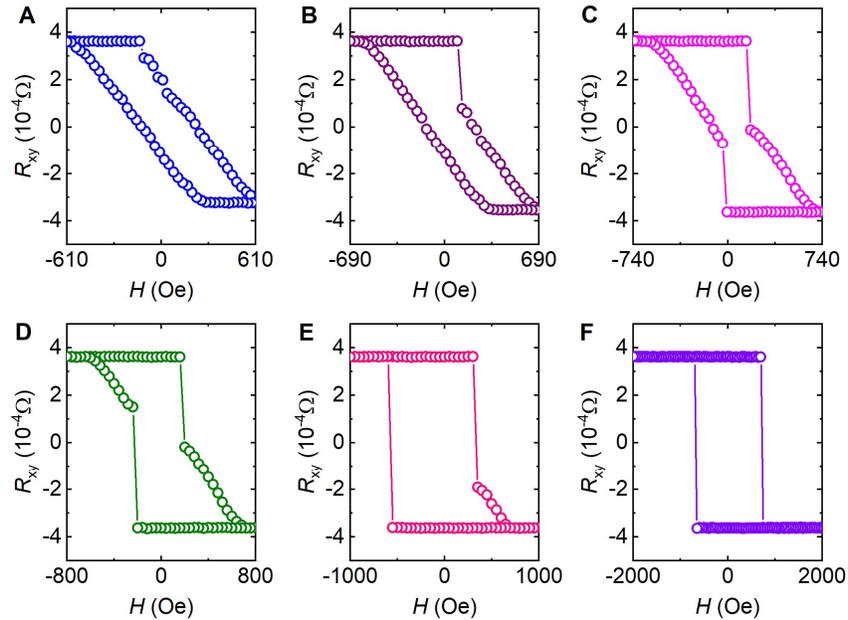

**Figure 5. Bow-tie-like Hall resistance hysteresis loops.** (**A**)-(**F**) Results with magnetic fields sweeping up to $H$ = 610 Oe, 690 Oe, 740 Oe, 800 Oe, 1 kOe and 2 kOe, respectively. The $R_{xy}$ ~ $H$ curves were taken with magnetic fields along the $c$-axis ($\theta$ = 180°) after driving the FM structure with an in-plane field up to $H_{ab}$ = 3.5 T. The FM structure was prepared by sweeping the field up to $H$ = -9 T along the $c$-axis and sweeping back to zero. Detailed procedures are presented in Figure S12 and its caption. Data was taken from Sample R3 at $T$ = 3 K.

## 4. Experimental Section/Methods

*Crystal Growth and Characterizations*: Single crystals of $Co_3Sn_2S_2$ were grown by flux method with tin as the flux. Stoichiometrical metal powders of Co (99.999%), Sn (99.999%), and S (99.95%) were mixed thoroughly in an alumina crucible, and then sealed into an evacuated quartz tube. The mixture was first heated to 1050°C and kept at that temperature for 10 hours. After that, the mixture was slowly cooled down to 740°C over 5 days, and single crystals were separated from the Sn flux by centrifugating. The inset of Figure S2 shows an image of a millimeter-sized crystal. The crystal structure was characterized by an X-ray diffractometer (Panalytical Empyrean), with a typical XRD pattern presented in Figure S2. The stoichiometry of the components was confirmed by using the Energy Dispersive Spectrometer (BRUKER EDS) equipped in a scanning electron microscope (EVO MA10).

*Transport measurement*s: We conducted resistance measurements on four crystals (hereby named Samples R1-R4) using the Electrical Transport Option (ETO) of Quantum Design PPMS®, which show consistent results. Electrical leads were gold wires glued to the crystal using silver epoxy H20E. A low-frequency (21.36 Hz) ac current of 1 mA was used and flows in the *ab* plane. Angular dependencies of the resistance were obtained by placing the sample on a precision, stepper-controlled rotator with an angular resolution of 0.05°. The magnetic field can rotate in a plane perpendicular to the current (Samples R1, R3 and R4) or in the plane determined by the *c*-axis and the current direction (Sample R2). The magnetic field orientation is represented by $\theta$, which is defined as the angle between the positive direction of the *c*-axis and the magnetic field vector pointing in the positive direction of the magnetic field, as shown in Figure 4A. More information on transport measurement procedures can be found in Figures S3 and S6 as well as their captions.

*Magnetization measurements*: Magnetization measurements were carried out in a Quantum Design MPMS®3 on two crystals (hereby named Samples M1 and M2). The change of the magnetic field orientation relative to the c-axis of the crystal is realized by placing the sample on a horizontal rotator (Model 301), which has an angular step size < 0.1°. The definition of θ is the same as that in transport measurements, i.e., θ = 0° (180°) for the magnetic field along the positive (negative) direction of the c-axis and θ = 90° for the field being parallel to the ab plane. More information on magnetization measurement procedures is presented in Figure S8 and its caption.

**Supporting Information**

Supporting Information is available from the Wiley Online Library or from the author.

**Acknowledgements**

We thank Shizeng Lin, Igor V. Solovyev and Vitalii Vlasko-Vlasov for stimulating discussions. Experimental design, transport and magnetization measurements were supported by the U.S. Department of Energy, Office of Science, Basic Energy Sciences, Materials Sciences and Engineering. S.E.P & Z.L.X acknowledge support from the National Science Foundation grant# DMR-1901843. Y. Z. was supported by the National Science Foundation Materials Research Science and Engineering Center program through the UT Knoxville Center for Advanced Materials and Manufacturing (DMR-2309083). B.W. & B.S. were supported by the National Key R&D Program of China (Grant No.2023YFF0718400), National Natural Sciences Foundation of China (Grants No.U2130101) and Guangzhou Basic and Applied Basic Research Foundation (Grant No. 2023B151520013, 2022A151501003).

Supporting Information

**Tunale Anomalous Hall Effect in a Kagomé Ferromagnetic Weyl Semimetal**

*Samuel E. Pate, Bin Wang, Yang Zhang, Bing Shen, Enke Liu, Ivar Martin, J. Samuel Jiang, Xiuquan Zhou, Duck Young Chung, Mercouri G. Kanatzidis, Ulrich Welp, Wai-Kwong Kwok, and Zhi-Li Xiao\**

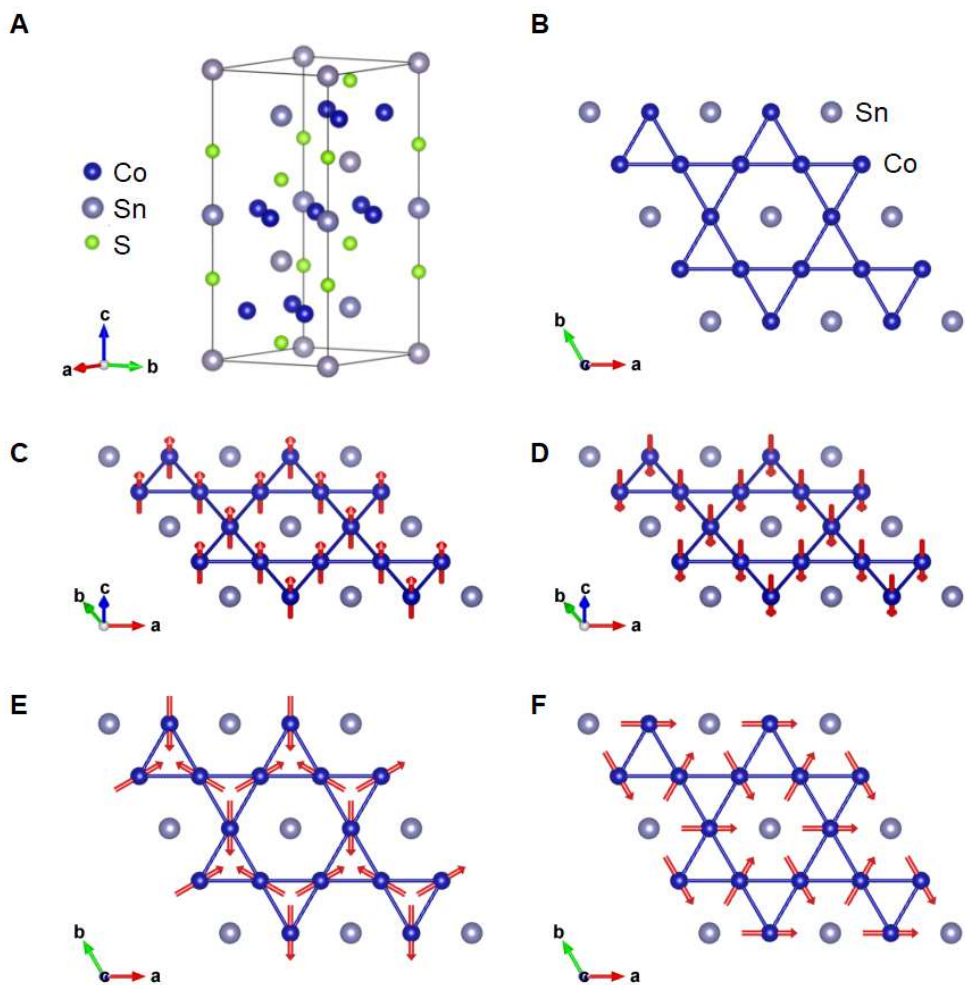

**Figure S1. Crystal and magnetic structures of Co₃Sn₂S₂.** ($\mathbf{A}$) Unit cell. ($\mathbf{B}$) Top view of the $Co_3Sn$ plane. ($\mathbf{C}$) and ($\mathbf{D}$) Ferromagnetic orders with spins along the positive and negative directions of the $c$-axis, respectively. ($\mathbf{E}$) and ($\mathbf{F}$) Anti-ferromagnetic structures with $R\bar{3}m'$ (E) and $R\bar{3}m$ (F) symmetries, respectively. Images were produced using VESTA opensource crystal modeling software.[11] Lattice parameters were taken from mp-19807: $Co_3(SnS)_2$ (Trigonal, $R\bar{3}m$, 166) (materialsproject.org).

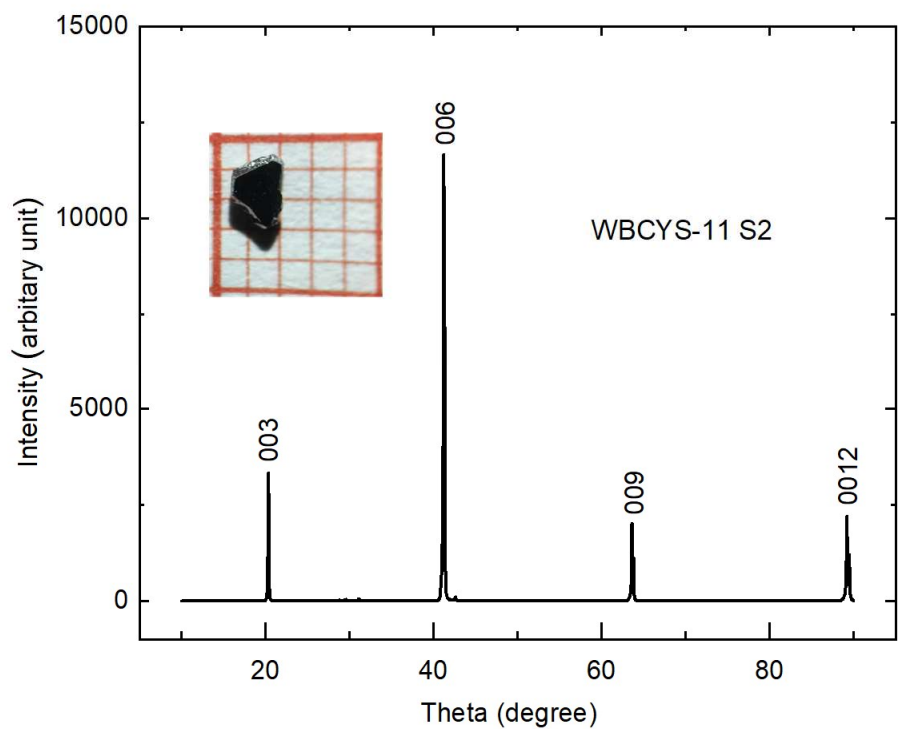

**Figure S2. X-ray diffraction pattern of a Co₃Sn₂S₂ crystal.** The peaks can be indexed by the indices of (00*l*) lattice planes. Inset presents a photo of a typical single crystal on a 1 mm grid paper.

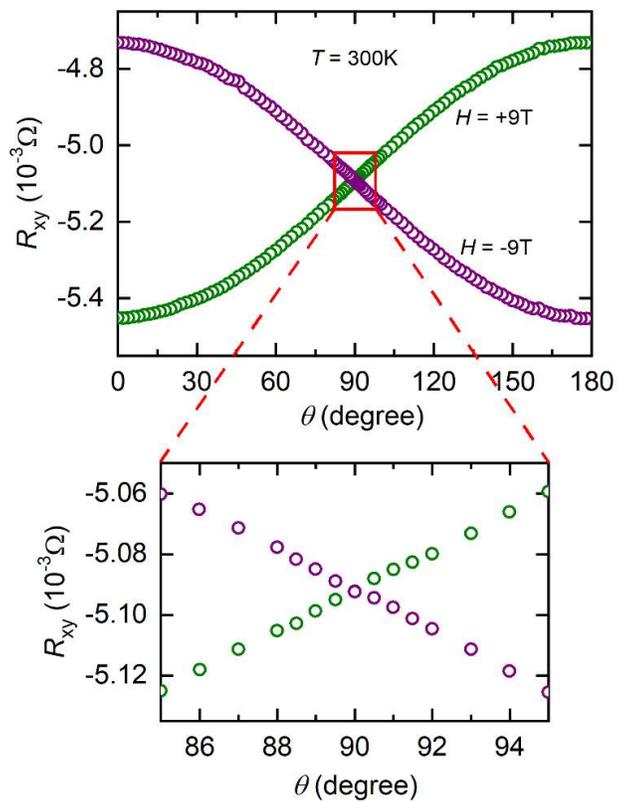

**Figure S3. Determination of the kagomé (*ab*) plane.** Hall resistance versus angle ($R_{xy} \sim \theta$) curves were taken at $H$ = +9 T and -9 T. The angle at which the two curves cross each other is defined as $\theta$ = 90°, corresponding to the kagomé (*ab*) plane. Data were taken from Sample R2 at $T$ = 300 K.

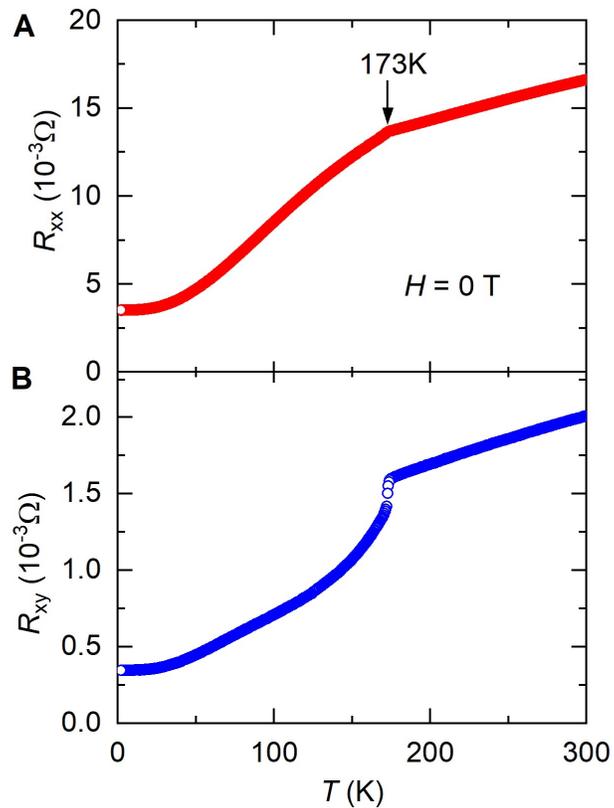

**Figure S4. Cooling resistance curves of a Co₃Sn₂S₂ crystal.** (**A**) and (**B**) present the longitudinal resistance $R_{xx}$ and the Hall resistance $R_{xy}$, respectively. Data were taken from Sample R2 at zero field. The Curie temperature of $T_c \approx 173$ K is indicated in (A).

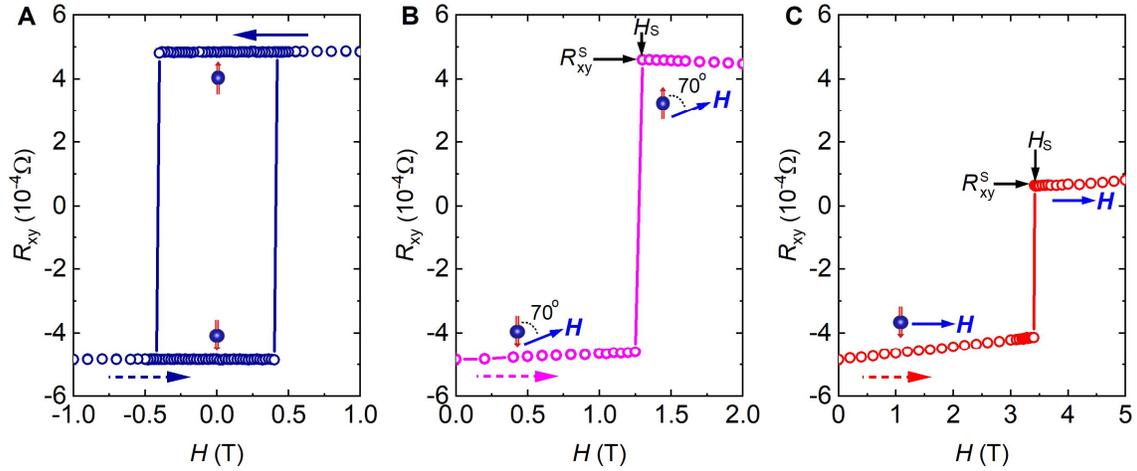

**Figure S5. Hall responses of Sample R2.** (**A**) Hysteresis loop of the Hall resistance versus the magnetic field at $H//c$. The magnetic field was swept up to $H = \pm 9$ T. (**B**) and (**C**) Hall responses of an FM order to magnetic fields orientated at $\theta = 70°$ and $90°$ (in-plane), respectively. The FM order was prepared by sweeping the field up to $H = -9$T along the $c$-axis and sweeping back to zero. The magnetic field rotates in the plane defined by the current direction and the $c$-axis, with $\theta$ being the angle between the positive direction of the $c$-axis and the field vector pointing in the positive direction of the magnetic field (Figure 4A). $\mathbf{H_s}$ and $\mathbf{R_{xy}^S}$ in (B) and (C) are the field and Hall resistance right after the spin switching. Data was taken at $T = 3$ K.

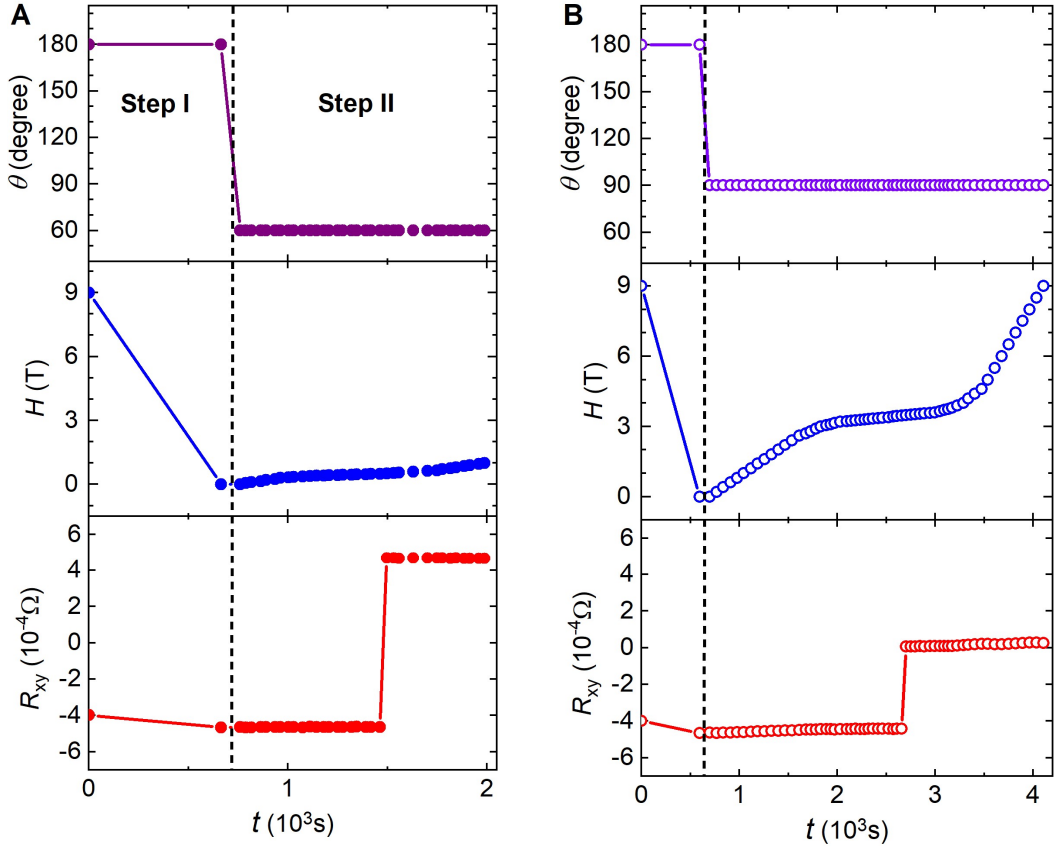

**Figure S6. Procedures for preparing and driving the FM order in the $R_{xy}$ measurements.** (**A**) and (**B**) Time evolution of the angle ($\theta$), magnetic field ($H$) and Hall resistance ($R_{xy}$) for the driving magnetic field at $\theta = 60°$ and $90°$, respectively. The FM order was first prepared by sweeping the field up to $H = 9$ T along the negative direction of the $c$-axis, i.e., at $\theta = 180°$ and sweeping back to zero, as marked as Step I in (A). It was then driven by a magnetic field orientated at various directions, e.g., $\theta = 60°$ in (A) and $\theta = 90°$ in (B). Dense data points of $R_{xy}$ were taken when the field sweeps up, as shown in the regime marked as Step II in (A). The slope change in the $H$ versus time curves in Step II is due to the change of field intervals, since smaller intervals around $H_s$ help to determine its value more accurately. The dashed lines in (A) and (B) separate Step I and Step II. The corresponding $R_{xy} \sim H$ data are plotted in Figures 1E and 1F, respectively.

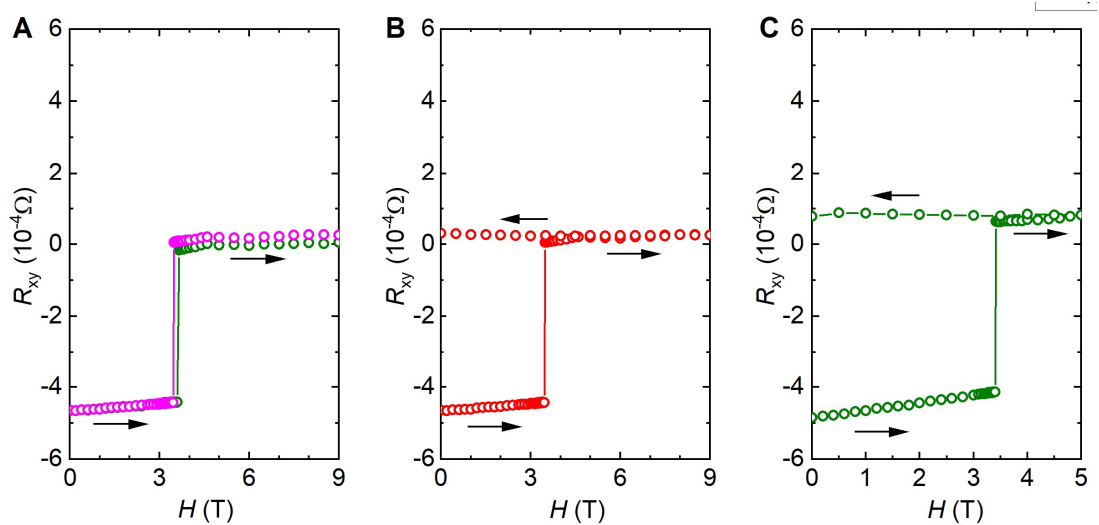

**Figure S7. Repeatability of the measurements and stability of the states after the sudden change in $R_{xy}$.** (**A**) Comparison of two sets of measurements of the same procedure (Figure S6) on Sample R1, demonstrating the repeatability. (**B**) and (**C**) Field excursion effects on the $R_{xy}$ after the sudden change for Sample R1 up to $H = 9$ T and for Sample R2 up to $H = 5$ T and then back to zero. exhibiting the stability of the states.

Following the procedure in Figure S6, the FM order was first prepared by sweeping the field up to $H = 9$ T along the negative direction of the *c*-axis (i.e., at $\theta = 180°$ ) and sweeping back to zero. It was then driven by an in-plane field. That is, the data shown in (A)-(C) were taken by sweeping magnetic field at $\theta = 90°$. All data were taken at $T = 3$ K.

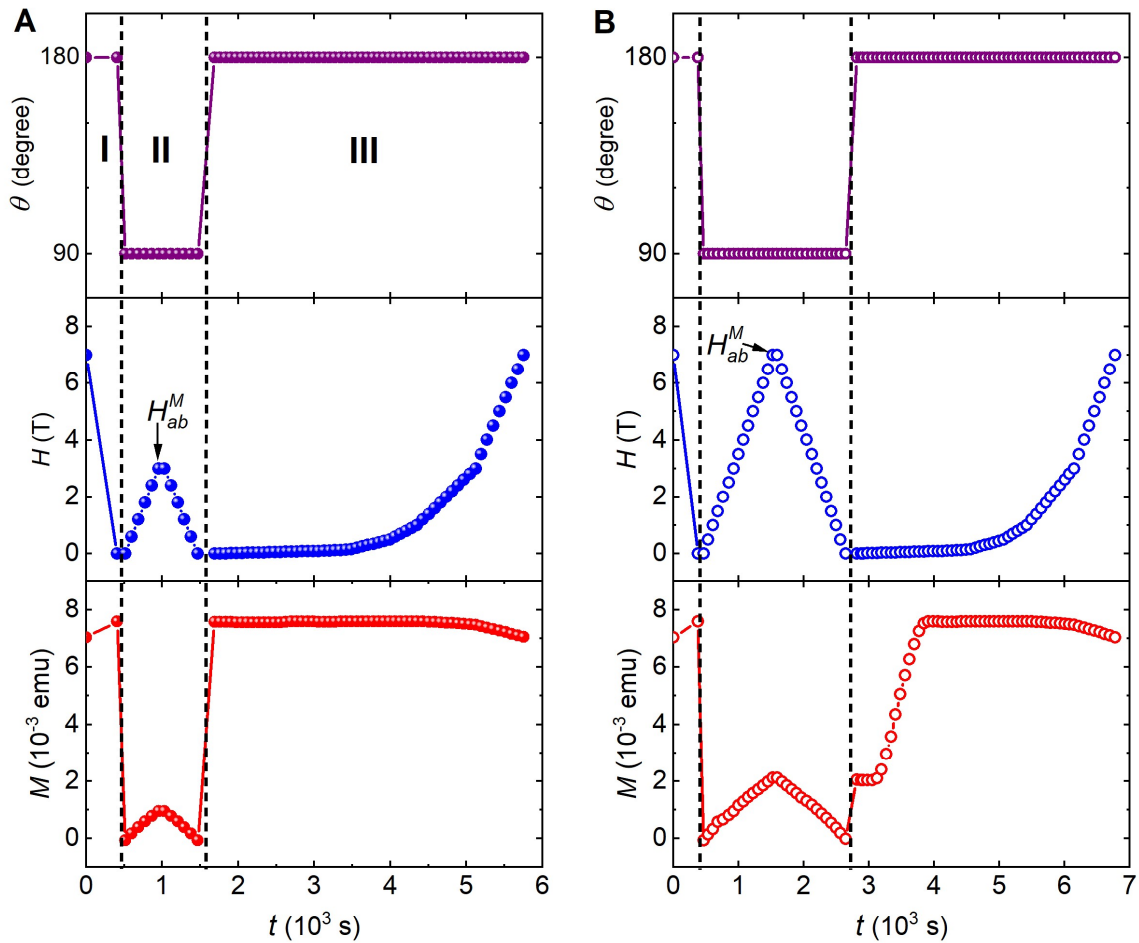

**Figure S8. Magnetization measurement procedures for detecting the magnetic states of the FM structure after being driven by an in-plane field.** (**A**) and (**B**) For in-plane fields up to $H_{ab}^{M}$ = 3 T and 7 T, respectively. The FM structure was first prepared by sweeping the field up to $H$ = 7 T along the negative direction of the $c$-axis, i.e., at $\theta$ = 180° and sweeping back to zero, as marked as Step I in (A). It was then driven by an in-plane magnetic field, i.e., at $\theta$ = 90°, up to various to $H_{ab}^{M}$ and the back to zero field, as marked as Step II in (A). After that, the sample is rotated back to $\theta$ = 180° and magnetization was measured from $H$ = 0 T to 7 T, as shown in the regime marked as Step III in (A). The change of the slope of the $H$ versus time curves in Step III is due to the variation of the used field intervals. The dashed lines in (A) and (B) separate the three steps. The corresponding $M_{ab} \sim H$ curves (from Step II) and $M_c \sim H$ curves (from Step III) are plotted in the insets and main panels of Figures 2A and 2B, respectively.

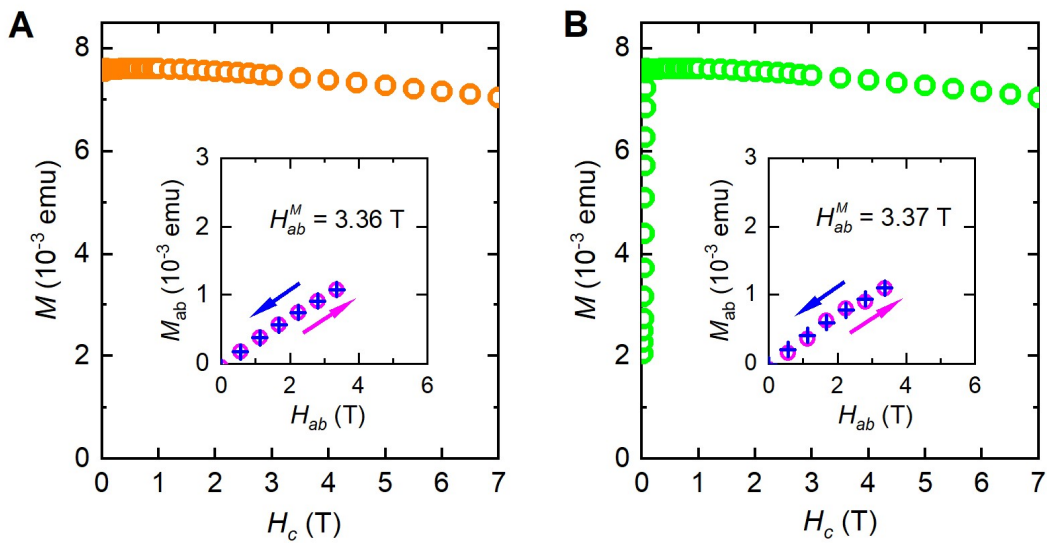

**Figure S9. Magnetization of the FM order being and after driven by an in-plane field.** (**A**) and (**B**) *C*-axis magnetization $M_c$ versus magnetic field $H_c$ curves of the FM order after driven by an in-plane field $H_{ab}$ up to $H_{ab}^M$ = 3.36 T and 3.37 T, respectively. Their insets present the in-plane magnetization $M_{ab}$ for the FM order driven by the in-plane field $H_{ab}$. The FM order was prepared by sweeping the field up to $H$ = 7 T along the *c*-axis and sweeping back to zero. Detailed procedures are presented in Figure S8 and its caption. Data were taken from Sample M1 at *T* = 3 K.

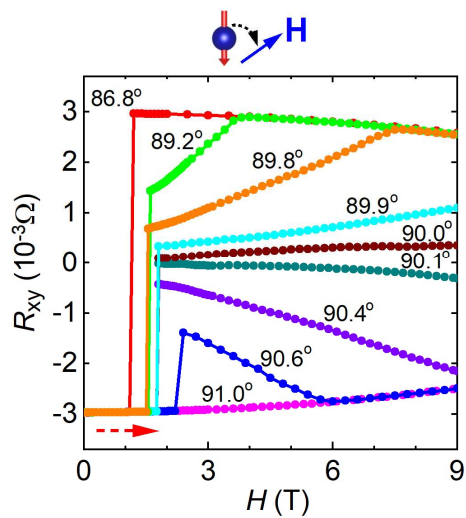

**Figure S10. Field-driven FM instability at a high temperature.** The FM order with spins pointing to the negative direction of the *c*-axis was prepared by sweeping the field up to *H* = - 9 T along the *c*-axis and sweeping back to zero. The red dashed arrow indicates the sweeping directions of the magnetic field for measuring $R_{xy}$. Numbers represent the angles at which the $R_{xy} \sim H$ curves were taken. The schematic on top of the main panel shows the directions of the spins in the FM order and the magnetic field as well as the definition of the angle. The magnetic field rotates in the plane perpendicular to the current direction. Data were taken from Sample R3 at *T* = 80 K.

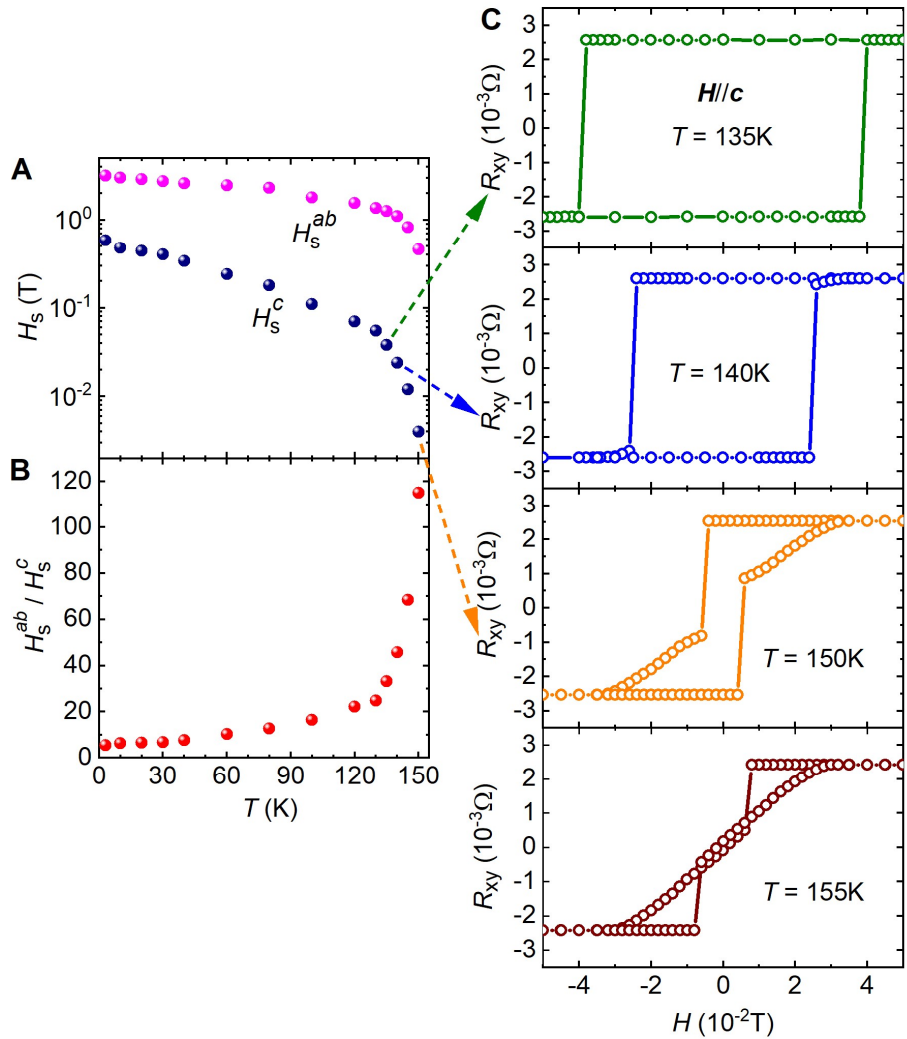

**Figure S11. Temperature effects on the local FM instability and bow-tie-like hysteresis loops.** (**A**) Temperature dependence of the critical fields $H_s^{ab}$ and $H_s^c$ for the local FM instability for the magnetic fields along the *ab* plane and the *c*-axis, respectively. (**B**) The temperature dependence of their ratio $H_s^{ab}/H_s^c$. (**C**) $R_{xy} \sim H$ loops at high temperatures, showing bow-tie-like features. The data were taken from Sample R4.

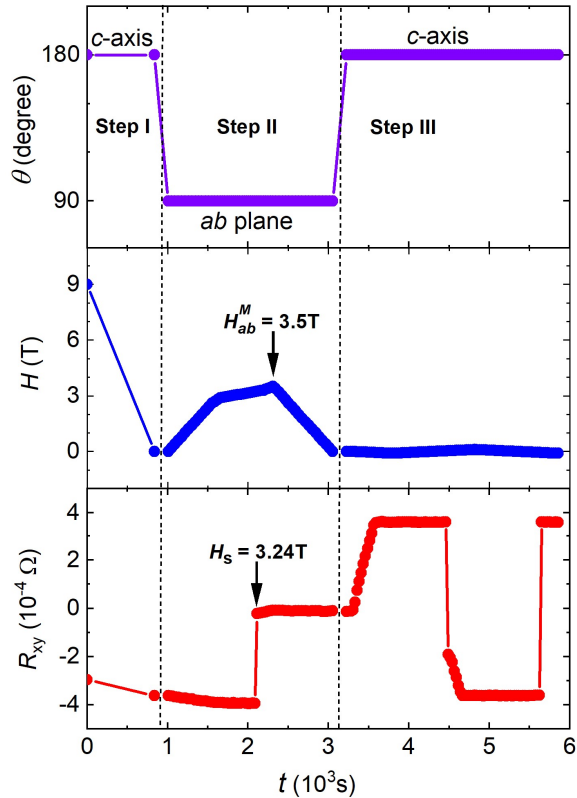

**Figure S12. Transport measurement procedures for detecting the magnetic states of the FM structure after being driven by an in-plane field.** The FM structure was first prepared by sweeping the field up to $H = 9$ T along the negative direction of the $c$-axis, i.e., at $\theta = 180°$ and sweeping back to zero (marked as Step I). It was then driven by an in-plane magnetic field, i.e., at $\theta = 90°$, up to various to $H_{ab}^M = 3.5$ T which is larger than $H_s = 3.24$ T, and then back to zero field (marked as Step II). After that, the sample is rotated back to $\theta = 180°$ at which $R_{xy}$ was measured from $H = 0$ T to $H = -0.1$ T to $H = 0.1$T and back to $H = -0.1$T, as shown in the regime marked as Step III. The $R_{xy} \sim H$ loop from $H = -0.1$ T to $H = 0.1$ T and back to $H = -0.1$ T is presented in Figure 5E. The data were taken from Sample R4 at $T = 3$ K.

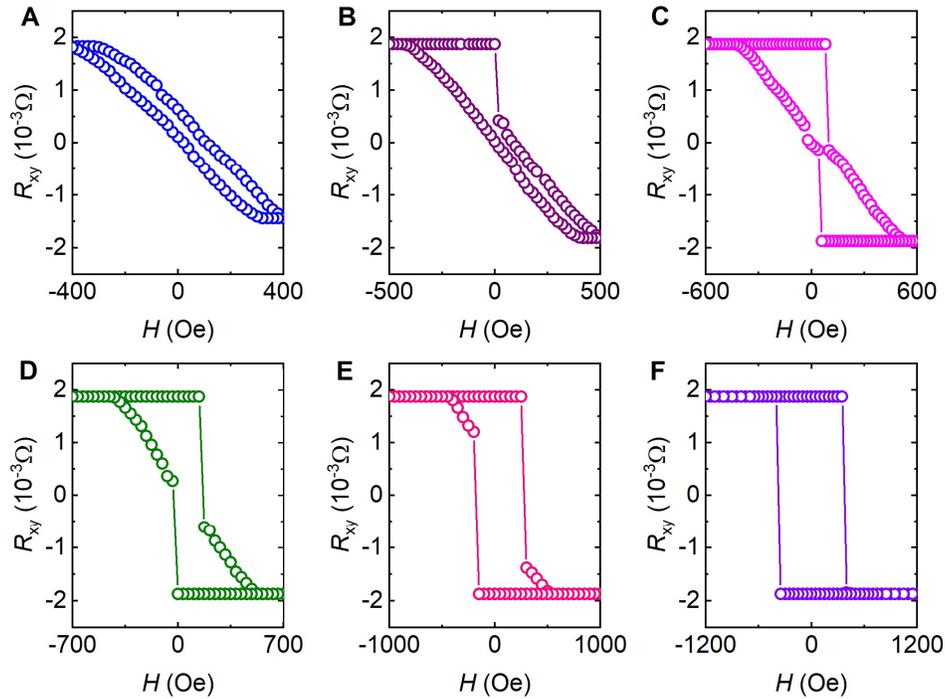

**Figure S13. Bow-tie-like Hall resistance hysteresis loops.** (**A**)-(**F**) Results with magnetic fields sweeping up to $H = \pm 400$ Oe, $\pm 500$ Oe, $\pm 600$ Oe, $\pm 700$ Oe, $\pm 1$ kOe and $\pm 1.2$ kOe, respectively. The $R_{xy} \sim H$ curves were taken with magnetic fields along the $c$-axis ($\theta = 180°$) after the FM structure had been driven with an in-plane field up to $H_{ab} = 2.2$ T. The FM structure was prepared by sweeping the field up to $H = -9$ T along the $c$-axis and sweeping back to zero. Data were taken from Sample R4 at $T = 100$ K. Measurement procedures are the same as those for $T = 3$ K presented in Figure S12 and its caption.